%% file: harmonization_paper.tex
\documentclass[runningheads]{llncs}
\usepackage{graphicx}
\usepackage{color}
\usepackage{amsmath}
\usepackage{hyperref}

\begin{document}

\title{Relevance Vector Machines for harmonization of MRI brain volumes using image descriptors 
\thanks{This project received funding from the European Union's Horizon 2020 research and innovation program under the Marie Sklodowska-Curie grant agreement No 765148.
The final authenticated publication is available online at \url{https://doi.org/10.1007/978-3-030-32695-1_9}
}}

\titlerunning{ RVM for harmonization of brain volumes using on image descriptors.}

\author{Maria Ines Meyer\inst{1,2}\and
Ezequiel de la Rosa\inst{1} \and \\
Koen Van Leemput \inst{2,3} \and Diana M. Sima \inst{1}}

\authorrunning{M. I. Meyer, E. de la Rosa, K. Van Leemput and D. Sima}
%
\institute{icometrix, Leuven, Belgium  \\
\email{\{ines.meyer, ezequiel.delarosa, diana.sima\}@icometrix.com}
 \and Dept. of Health Technology, Technical University of Denmark, Lyngby, Denmark \\ 
 \and Martinos Center for Biomedical Imaging, Massachusetts General Hospital \& Harvard Medical School, Boston, MA \\ 
\email{{kvle}@dtu.dk}}


\maketitle             
\begin{abstract}
With the increased need for multi-center magnetic resonance imaging studies, problems arise related to differences in hardware and software between centers. Namely, current algorithms for brain volume quantification are unreliable for the longitudinal assessment of volume changes in this type of setting. Currently most methods attempt to decrease this issue by regressing the scanner- and/or center-effects from the original data.
In this work, we explore a novel approach to harmonize brain volume measurements by  using only image descriptors. First, we explore the relationships between volumes and image descriptors. Then, we train a Relevance Vector Machine (RVM) model over a large multi-site dataset of healthy subjects to perform volume harmonization.
Finally, we validate the method over two different datasets: i) a subset of unseen healthy controls; and ii) a test-retest dataset of multiple sclerosis (MS) patients. The method decreases scanner and center variability while preserving measurements that did not require correction in MS patient data. We show that image descriptors can be used as input to a machine learning algorithm to improve the reliability of longitudinal volumetric studies.
\end{abstract}

\input{intro.tex}

\input{methods.tex}
\input{results.tex}
\input{discussion.tex}

%
%
%
%
\bibliographystyle{splncs04}
\bibliography{references}

\end{document}

%% file: intro.tex
\section{Introduction}

\begin{figure}[t]
    \setlength{\belowcaptionskip}{-10pt}
    \centering
    \includegraphics[width=\textwidth]{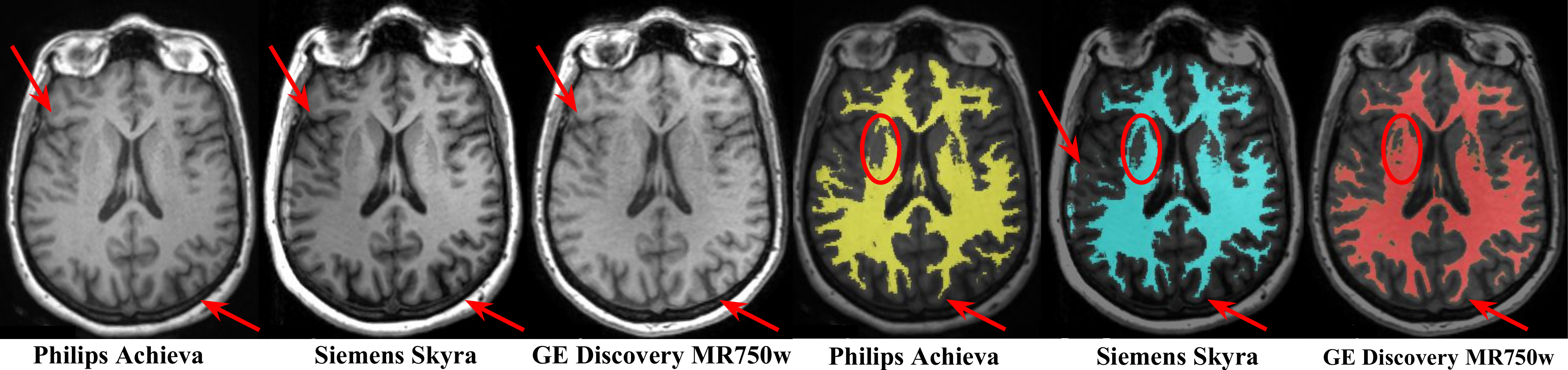}
    \caption{MR images from same patient in different scanners. \textbf{Left side:} T1-weighted images. \textbf{Right side:} white matter segmentations obtained using the same method. }
    \label{fig:scan_comparison}
\end{figure}

Large scale multi-site studies are of extreme importance in neuroimaging, both for research purposes and in clinical practice. Such studies face several challenges due to hardware- or center-related variability.  
It is well known that scanner-factors such as manufacturer, magnetic field and gradient non-linearly influence volume measurements \cite{Chen2014,Takao2011} obtained from structural Magnetic Resonance Imaging (MRI). At the image level, these factors are coupled with a high variability of intensities across patients and scanners, which can affect tasks like the segmentation of brain structures \cite{Zhuge2009}. This effect is exemplified in Fig. \ref{fig:scan_comparison}, where three T1-weighted MR images from the same patient obtained on different scanners and their corresponding segmentations are represented. 

The need to address multi-scanner and -center data harmonization is evidenced in the follow-up of Multiple Sclerosis (MS) patients. These patients exhibit an increased rate of brain atrophy when compared to healthy subjects, which has been linked to impairment \cite{Bermel2006}. However, it has been suggested that brain atrophy can only be reliably estimated over periods of at least five years \cite{Biberacher2016}, due to the variability caused by scanner and center factors.  

Besides image processing approaches that aim at matching image intensity distributions to provide a more consistent input to the segmentation method \cite{Nyu1999,Robitaille2012}, recent studies have focused on statistical harmonization of volumetric measurements based on scanner- or center-specific information. 
This type of methods generally apply regression techniques to correct measurements. 
Linear mixed-effects models using patient- and scanner-specific information as random effects were explored by \cite{Chua2015} and \cite{Jones2013}. Recently, \cite{fortin2018_combat} used an algorithm devised for genomics that extends the same type of model to account for site-specific factors. A data-driven approach based on independent component analysis was explored by \cite{Chen2014}, where correction was performed by selecting independent components related to scanning parameters. However, since these methods rely on scanner- or acquisition-specific information, they do not generalize and need to be adapted when used in new settings.
Additionally, such information can be incomplete, especially in historical data. 
As such, it would be of interest to use information that is encoded in the images themselves, or that can be extracted from the volume quantification method to build more robust and adaptive techniques.  

To address these issues, in this paper we present a novel statistical harmonization approach based on image descriptors and a machine learning algorithm. We first explore the relations between image-extracted properties and brain volume measurements that we could further exploit for harmonization. We then train a machine learning algorithm based on Automatic Relevance Determination on healthy data to perform volumetric corrections. We validate the method on a set of unseen healthy controls, and finally test it on a a test-retest dataset of MS patients.

%% file: methods.tex
\section{Data}
\label{sec:data}
\paragraph{\textbf{Healthy Subjects}} 
This dataset comprises 1996 T1-weighted (T1w) MRI scans from healthy subjects. The data is a compilation of several public datasets, such as  \cite{OASIS_cross,ixi_online}, and some proprietary data.
The overall set comprises data from several different centers and scanner types from the major vendors (Siemens, Philips, GE). Magnetic field strengths (1.5T or 3T) and T1w sequence types also vary.
For most of the data we have information regarding age and sex of the subject, scanner type, magnetic field strength and additional acquisition parameters like echo time (TE) and repetition time (TR). 
For building and testing our model, we randomly divided the data into training ($70 \%$) and test sets ($30 \%$).

\paragraph{\textbf{Patient Data}}  
To further validate the approach we test it in a dataset containing data from 10 MS patients as detailed in \cite{Jain2015}. Each patient was scanned twice in three different 3T scanners: \emph{Philips Achieva}, \emph{Siemens Skyra} and \emph{GE Discovery MR450w}. An example is depicted in Fig.  \ref{fig:scan_comparison}. 
We observed that one of the patients was an extreme case, showing very enlarged ventricles. Given that the volumetric measurements in such a case are prone to errors and are considered unreliable, this patient's data was discarded from further analysis. 

\subsection{Data Pre-processing and Feature Extraction}
For each image we compute gray matter (GM) and white matter (WM) volumes using the well established atlas-based method described in {\cite{Jain2015}}. Whole brain (WB) volume is then defined as the sum of WM and GM volumes. 

We are interested in descriptors related to the T1w images that encode information about errors and bias in brain segmentations. Since the quality of a segmentation depends on a good registration to the atlas and is influenced by the contrast and noise present in an image, it is valuable to explore features that convey such information. 
We extract a total of 16 features of two main types: i) Alignment information regarding the registration of the T1w image to the MNI atlas space, which includes decomposing the affine transformation (rotation angles, scale and shear factors in three directions), and measuring the similarity between the registered images using Normalized Mutual Information (NMI); and ii) Contrast to Noise Ratio (CNR) between tissue types and between different brain structures (\emph{e.g.}, lobes and cerebellum). CNR is given by:  
\begin{center}
$
CNR_{t_1,t_2} =   
\sqrt{2} {\frac {{|\bar{I}_{t_1} - \bar{I}_{t_2 }|}}
         {\sqrt{\sigma^{2}_{t_1} + \sigma^{2}_{t_2}}}},
$
\end{center}
where $\bar{I_t}$ represents the mean intensity of some tissue or structure $t$ and $\sigma^2_t$ is the variance of the image intensities across this structure. We compute $CNR_{t_1, t_2}$ by taking $t_1$ as the tissue or structure with higher average image intensity than $t_2$. 

The brain volumes and some of the computed descriptors (\emph{e.g.}, CNR and NMI) are known to be age dependent \cite{Salat2009}. As such, age is used as a feature at training time, but not at test time. For analysis and comparison, we age-detrend the volumes by subtracting an age-matched estimated median value. CNR and NMI are corrected by fitting a linear regressor to the data. 

\section{The Relevance Vector Machine for data harmonization} \label{sec:methods-rvm}
To harmonize brain volumes, we subtract correction terms based on estimated variability trends from the original volumes. To determine the variation in the volumetric data that can be explained by the aforementioned image descriptors, we fit a linear model using the extracted features as independent variables. 
When choosing the model, we take a few important considerations into account: i) we have 16 image-extracted features plus age; ii) some of these are related only to one of the brain volumes; iii) the features are not always unrelated; and iv) we are interested in a probabilistic model, to capture uncertainty in our predictions. 
Given that standard generalized linear models do not address all these considerations, we investigate using a probabilistic machine learning technique. 

The \emph{Relevance Vector Machine} (RVM) is defined within a fully probabilistic framework and includes a mechanism of \emph{automatic relevance determination} \cite{MacKay1992}. 
As described in \cite{Tipping_rvm_2001}, the model defines a conditional distribution  for real-valued input-target vector pairs $\{\mathbf{x}_n, t_n\}_{n=1}^{N}$, of the type: 
$p(t_n|\mathbf{x}_n) = \mathcal{N}(t_n|y(\mathbf{x}_n), \beta^{-1}),$ 
which specifies a Gaussian distribution over $t_n$ with mean $y(\mathbf{x}_n)$ and precision (inverse variance) $ \beta$. Here $\{\mathbf{x}_n\}_{n=1}^{N}$ are the set of extracted features and $\{t_n\}_{n=1}^{N}$ the corresponding volume measurement for each image $n$ in a training dataset of size $N$.
The function $y(\mathbf{x})$ is given by a linear combination of basis functions 
$\boldsymbol{\Phi}(\mathbf{x}) = (\Phi_1(\mathbf{x}), \ldots, \Phi_N(\mathbf{x}))$
with weights $\mathbf{w} = (w_0, \ldots, w_N)^{\mathrm{T}}$:
\begin{center}
$ y(\mathbf{x}) = \sum_{i=1}^{N}w_i \Phi_i(\mathbf{x}) + w_0 = \boldsymbol{\Phi}(\mathbf{x}) \mathbf{w},$
\end{center}
where the $i^{\mathrm{th}}$ basis function $\Phi_i(\mathbf{x}) \equiv K(\mathbf{x},\mathbf{x_i})$ is a kernel centered around the $i^{\mathrm{th}}$ training sample. 
In order to avoid over-fitting due to the large numbers of parameters in the model, a zero-mean Gaussian prior probability distribution is defined over the weights $\mathbf{w}$. Moreover, a separate hyperparameter $\alpha_i$ is introduced for each individual weight  $w_i$, representing the precision of the corresponding weight: 
\begin{center}
$
p(\mathbf{w}|\boldsymbol{\alpha}) = \prod_{i=0}^{N} \mathcal{N}(w_i|0, \alpha_i^{-1}),
$
\end{center}
where $\boldsymbol{\alpha} = (\alpha_0, \ldots, \alpha_N)^{\mathrm{T}}$.
Using the resulting model, relevance vector learning searches for the hyperparameters $\boldsymbol{\alpha}$ and $\beta$ that maximize the marginal likelihood $p(\mathbf{t}|\boldsymbol{\alpha},\beta)$ of the training data, where $\mathbf{t} = (t_1, \ldots, t_N)^{\mathrm{T}}$.
Defining $\boldsymbol{\Phi}$ as the $N \times N $ matrix with  $\boldsymbol{\Phi}(\mathbf{x}_n)$ in $n^{\mathrm{th}}$ row,
the learning algorithm proceeds by iteratively updating $\boldsymbol{\alpha}$ and $\beta$ as follows~\cite{Tipping_rvm_2001}:
\begin{center}
$
\alpha_{i}^{new} = \frac{\gamma_i}{\mu_i^2} 
\quad
\mathrm{and}
\quad
(\beta^{new})^{-1} = \frac{\|\mathbf{t}-\boldsymbol{\Phi}\boldsymbol{\mu}\|^2}{N-\sum_i\gamma_i}  \quad\textrm{with} 
 \quad \gamma_i \equiv 1-\alpha_i\Sigma_{ii},  
$
\end{center} 
where $\boldsymbol{\Sigma} = \left(\beta \boldsymbol{\Phi}^{\mathrm{T}} \boldsymbol{\Phi} + \mathrm{diag}(\boldsymbol{\alpha})\right)^{-1}$ and $\boldsymbol{\mu} = \beta \boldsymbol{\Sigma} \boldsymbol{\Phi}^\mathrm{T} \mathbf{t}$ are the posterior covariance and mean for the weights, respectively. 
In practice, during re-estimation, many $\alpha_i$'s tend to infinity, which causes the posterior distributions of the corresponding weights to peak around zero. The basis functions associated with these do not influence the predictions and can be pruned out, resulting in a sparse model. The remaining training samples with non-zero weights are called \emph{relevance} vectors.

Once the model is trained, we can take a set of descriptors $\mathbf{x}_*$ of an unseen image, and try to predict the corresponding volume based on these descriptors alone using the posterior mean $\boldsymbol{\mu}$:  $y_* =  \phi(\mathbf{x}_*) \boldsymbol{\mu}$.
Finally, we can obtain a \emph{corrected} volume $y_{corr}$ by subtracting the estimated contribution of the image descriptors from the original volume $y$:
$ y_{corr} = y - \boldsymbol{\phi}(\mathbf{x}_*)\boldsymbol{\mu} $.

%% file: results.tex
\section{Results}
\subsection{Verification of observable correlations in data}
\begin{figure}[t]
    \centering
    \includegraphics[width=0.89\textwidth]{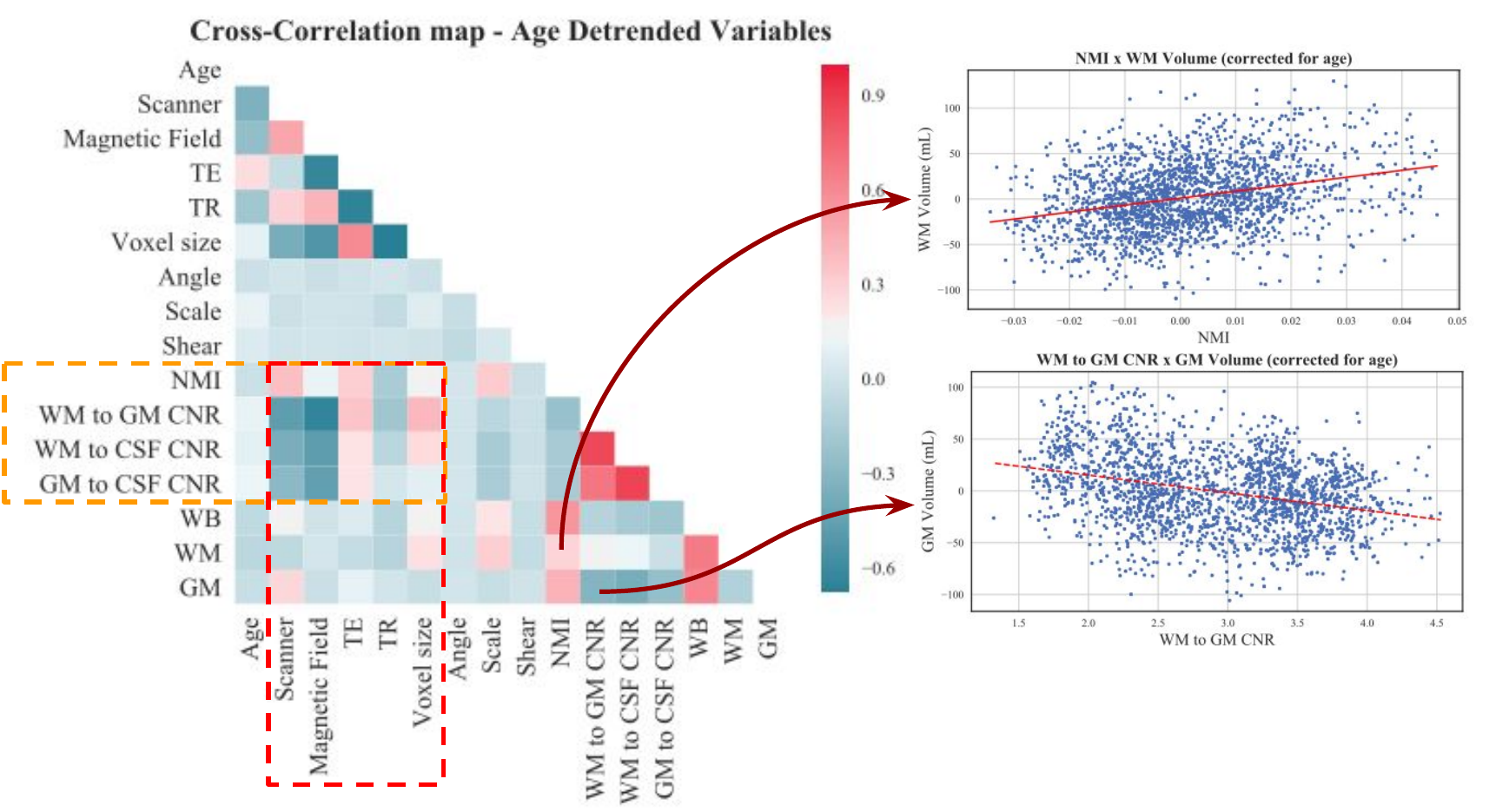}
    \caption{Left: Cross-correlations between age-detrended brain volumes (mL) and extracted features. We represent \emph{angle}, \emph{shear} and \emph{scale} as the result of multiplying the three directions.  Right: a zoomed-in view of the relationship between volumes and image descriptors: top WM vs.\ NMI and bottom GM vs.\ CNR.} 
    \label{fig:cross-corr}
\end{figure}

To verify whether there are correlations between image descriptors and the measured volumes, we built a cross-correlation map between these variables (see Fig. \ref{fig:cross-corr}). 
Analysing these correlations reveals that image descriptors like NMI and CNR are related to scanner/acquisition specific features. These same image descriptors are in turn correlated to the brain volumes, as well as scale.

\subsection{Harmonization of healthy population data based on RVM}
\begin{figure}[t]
    \centering
    \includegraphics[width=0.9\textwidth]{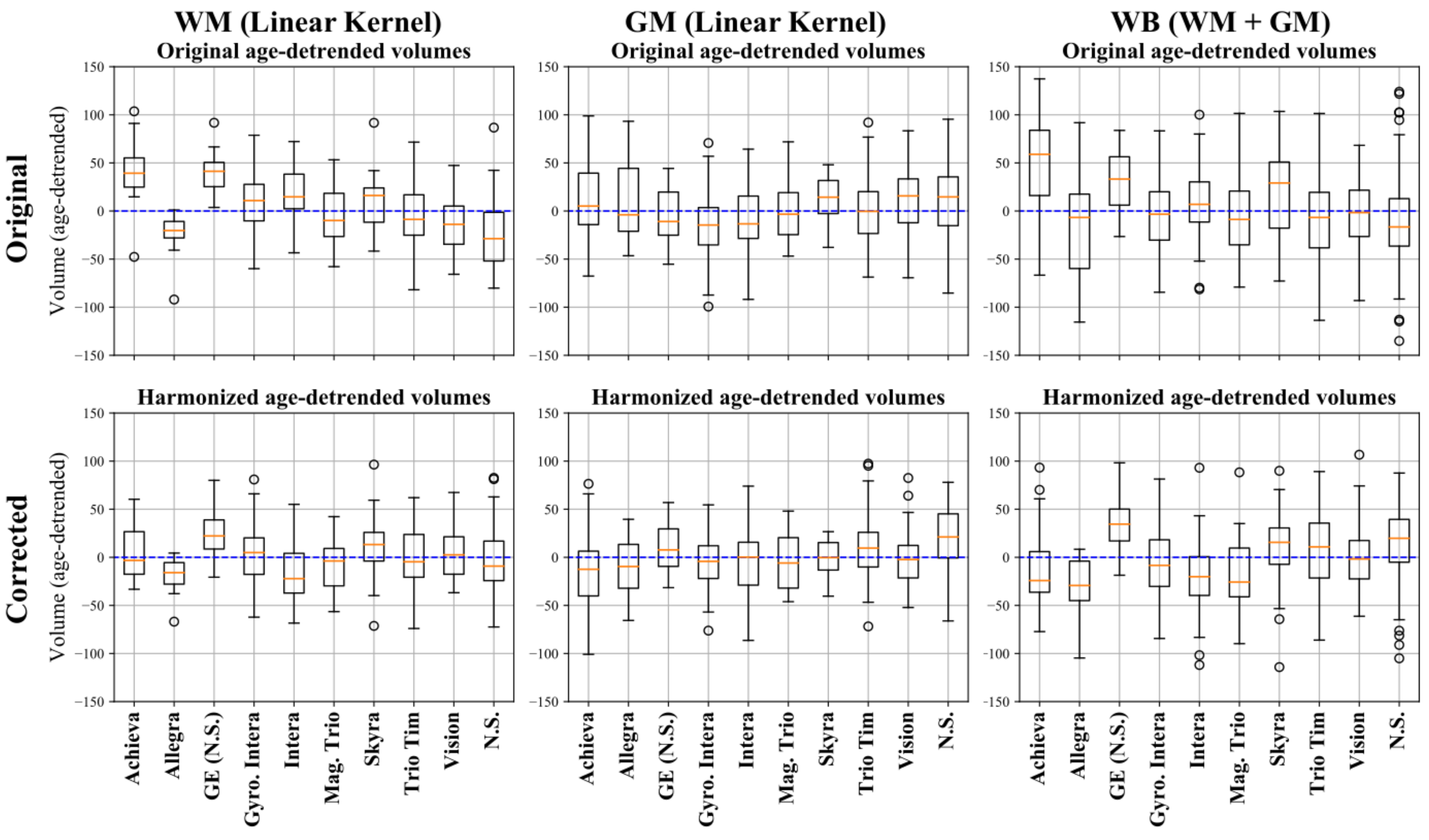}
    \caption{Distribution of the validation set volumes (mL) before (top) and after (bottom) RVM-based correction using a linear kernel. (N.S.: Not Specified).} 
    \label{fig:box-plots}
\end{figure}

We trained and tested the RVM method described in section \ref{sec:methods-rvm} for linear regression on the data set of healthy subjects (section \ref{sec:data}). We used different kernel types (linear and Gaussian - RBF) and searched for the model that best preserved biological information - namely age - while decreasing the scanner/center-specific variability. Thus, the model should decrease global variance in the data but maintain the original median of the population defined by the training set, given that we build on the assumption that this sample contains enough variability to represent the heterogeneity of scanner and center effects.
To evaluate the performance, we produced boxplots to represent the distribution of the measured volumes in each scanner in the test set with at least ten subjects (see Fig. \ref{fig:box-plots}). We first removed the age dependency as estimated from the training set, such that the variability due to age is not accounted for. We compare median and standard deviation, preferring values closer to zero, since they represent a decrease in variability while preserving the global trend. Table \ref{table:kernel_medians} presents the median values of these same age-detrended values. After correction the distributions from different scanners become more similar.
For both GM and WM the linear kernel produced lower or comparable mean and decreased standard deviation. For WB we compared applying a linear kernel to summing the previously corrected WM and GM volumes and verified that the last option performed better. 
\begin{table}[t]
\caption{Median and standard deviation of age-detrended volumes (mL) before and after applying the RVM-based correction with different kernels (linear and RBF). Data pertains to the scanner-wise distribution of the test set (as in Fig. \ref{fig:box-plots}). }

\centering{
\begin{tabular}[t]{ l | c c }
                        &   \multicolumn{2}{|c}{WM} \\ \hline
Kernel                  & Median   &   STD   \\
\hline
Original               &  10.9             &    30.4  \\
\textbf{Linear}        &  \textbf{-3.2}    &   \textbf{26.9} \\
RBF                    & 11.0              &   27.6\\ 
\end{tabular}
\begin{tabular}[t]{| c c |}
            \multicolumn{2}{|c|}{GM} \\\hline
             Median   &   STD        \\ 
\hline
             -0.3  &   32.5   \\
             \textbf{-0.3}  &   \textbf{29.0}     \\
              -7.5 &   33.5    \\ 
\end{tabular}
\begin{tabular}[t]{|l| c c} 
                        & \multicolumn{2}{|c}{WB} \\
\hline
Kernel       & Median   &   STD  \\
\hline
Original                & 1.2  &   47.9    \\
Linear                  & -3.4  & 38.2     \\
{\footnotesize WM+GM (linear)}          & \textbf{-0.8}  &   \textbf{37.3}    \\
\end{tabular}
}
\label{table:kernel_medians}
\end{table}

\subsection{Harmonization of test-retest data}
\begin{figure}[t]
    \centering
    \includegraphics[width=0.91\textwidth]{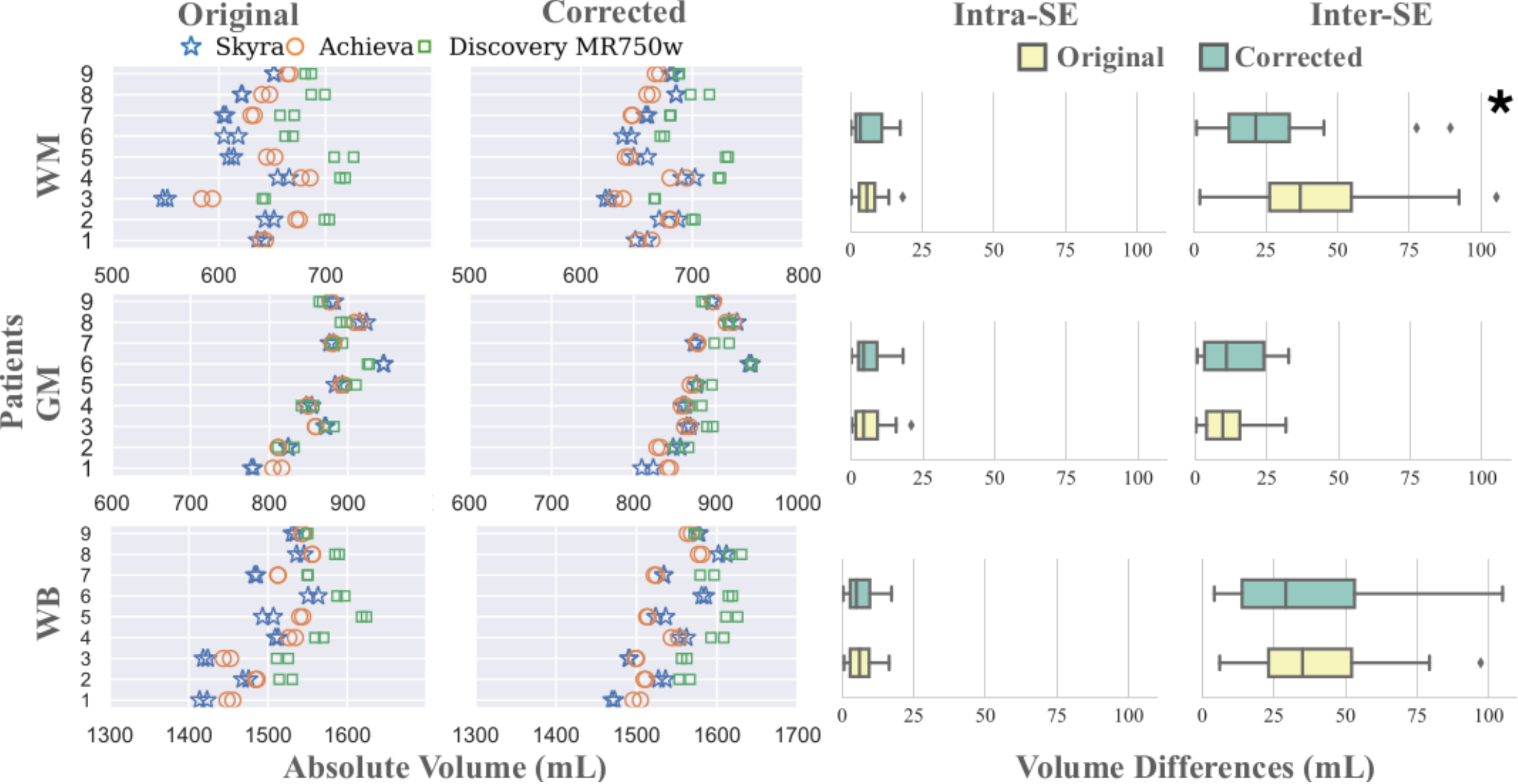}
    \caption{Difference between same patient brain volumes (mL) acquired in different scanner types. Left: Absolute volumes for each patient before and after correction. Right: Distribution of differences between the Intra-SE and Inter-SE volumes from same patient before and after correction. }
    \label{fig:re3t-plots}
\end{figure}

To further validate the method, we applied it to the test-retest dataset of MS patients described in Section \ref{sec:data}.
The results are summarized in Fig. \ref{fig:re3t-plots}. On the left side, the absolute volumes before and after correction are represented. First we computed for each tissue type the differences between the volumes from images acquired in the same scanner, which provides a measure of the intra-scanner error (Intra-SE). There is no significant difference between the original and corrected volumes for all the tissue types ($p>0.05$, paired t-test).
Then we computed the difference between the averaged volumes of each scanner type against all the other scanner types. For WM there is a statistically significant difference ($p=2e^{-3}$, paired t-test) between the original and corrected volumes. For GM, the inter-scanner error (Inter-SE) for the original volumes was very small, being comparable to the Intra-SE. After applying the correction, there is no statistical difference between these volumes and the original ones, even if visually there is an increase in the variability which is propagated for the WB. 

%% file: discussion.tex
\section{Conclusions and Future Work}
In this work we applied a relevance vector machine approach to find the amount of variability in the data that can be explained by variations in image descriptors.
We observe that there is a large dependency of brain volumes with the atlas-registered NMI metric, which was initially not expected. NMI measures the goodness of a non-rigid registration step between the image and an atlas, necessary for the methodology we used. The final volumes depend on the goodness of this registration, and in such a way we are correcting for suboptimal segmentation results that derive from a poor registration step. 

We demonstrate that it is possible to achieve a certain degree of harmonization of the data based only on image descriptors. 
To our knowledge, this is the first approach that does not rely on scanner-specific  information to perform harmonization. We expect the current method to perform less efficiently than more tailored methods, but to generalize better. A thorough comparison to such methods still needs to be performed, but it is out of the scope of the current paper. This type of solution is interesting for large scale statistics, and could potentially have a positive impact in longitudinal studies. Moreover, the proposed approach allows dealing with missing scanner/center information, a problem not addressed in previous works and very frequent in practice.
Nevertheless, in the test-retest setting inter-scanner error is still high when compared to the measured intra-scanner error, which implies that the method does not provide a completely satisfactory correction for patient specific use, and should be further investigated. 

Future steps include exploring more image descriptive features that are independent from the segmentation method used and that can encode the presence of geometrical distortions and artifacts. For controlled environments it could be useful to couple general scanner-dependent information with the image descriptors. Additionally, we aim to extend the method to other brain structures of interest and to compare its performance on a controlled dataset to scanner-specific state-of-the art methods.
Finally, it is important to keep in mind that in a cross-sectional setting this type of correction does not replace the need for an improved standardization at the image level.